# Modeling the fringe field of permanent magnet multipoles using numerical simulations

T. Thuillier [1, a)], T. André [1], and J.A. Mendez Giono [1]

[1]*Université Grenoble-Alpes, CNRS-IN2P3, Grenoble Institute of Engineering (INP), LPSC, 38000 Grenoble, France*

Permanent Magnet multipoles (PMM) are widely used in accelerators to either focus particle beams or confine plasma in ion sources. The real magnetic field created by PMM is calculated by magnetic field simulation software and then used in particle tracking codes by means of 3 dimensional magnetic field map. A common alternative is to use the so-called "hard edge" model, which gives an approximation of the magnetic field inside the PMM assuming a null fringe field. This work proposes an investigation of the PMM fringe field properties. An analytical model of PMM magnetic field is developed using the Fourier multipole expansion. A general axial potential function with a unique parameter λ, able to reproduce the actual PMM magnetic field (including its two fringe fields) with an explicit dependence on the PMM length is proposed. An analytical first order model including the axial fringe field is derived. This simple model complies with the Maxwell equations ($curl(B) = 0$ and $div(B) = 0$) and can replace advantageously the "hard edge" model when fast analytical calculation are required. Higher order analytical multiple expansion model quality is assessed by means of $\chi^2$ estimators. The general dependence of the potential function parameter λ is given as a function of the PMM geometry for quadrupole, hexapole and multipole, allowing to use the developed model in simulation programs where the multipole geometry is an input parameter.

## I. INTRODUCTION

Permanent magnets multipoles (PMM), as proposed by Halbach [1], are today widely used to focus particle beams in accelerators and to confine plasmas in ion sources. The PMM conveniently produces very high magnetic field gradients and high magnetic field intensities. When the multipole of order 2m with inner radius $R_0$ (see Fig.1, radius along the transverse directions $x$ and $y$) and length $L$ (along the $z$ direction) is sufficiently long ($L \gg R_0$), the inner magnetic field generated is usually approximated by a "hard edge" model with a pure transverse magnetic field whose components at point $M(x,y,z)$ are defined in cylindrical coordinates as:

$$\begin{cases} B_x = B_0 \left(\frac{r}{R_0}\right)^{m-1} \cos((m-1)\theta) \\ B_y = -B_0 \left(\frac{r}{R_0}\right)^{m-1} \sin((m-1)\theta) \\ B_z = 0 \end{cases} \quad (1)$$

where $B_0$ is a constant (*i.e.* intensity of the magnetic field at $r = R_0$), $r^2 = x^2 + y^2$ and $\theta = (\vec{x}, \vec{r})$. The magnetic field is assumed to be null outside of the multipole. This model is commonly used in simulations as it is easy to implement and allows for fast analytical calculation of magnetic field values (e.g. [2,3]).

a)Electronic mail: thuillier@lpsc.in2p3.fr

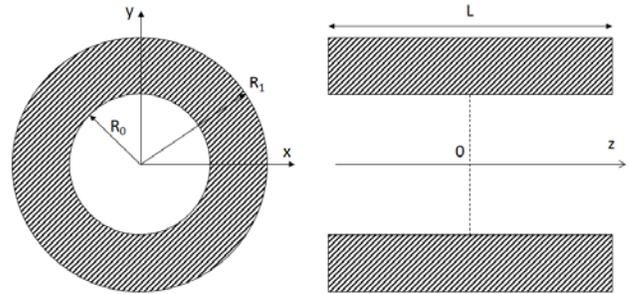

Fig. 1. Multipole geometrical parameters used. z is the multipole axis.

Nevertheless, in many cases the magnetic field close to the multipole edge plays a non-negligible role in the process studied (for instance ion beam formation from an electron cyclotron resonance ion source or particle beam focusing) and it becomes necessary to consider the magnetic fringe field. When considering a PMM with length $L$, inner and outer magnet radii respectively $R_0$ and $R_1$ (see Fig. 1), the magnetic fringe field expands on a characteristic lengh of the order of $R_0$. As an example, the real magnetic field line of a hexapole ($m = 3$) Halbach PMM with 36 PM segments and $R_0 = 50$ mm is plotted on Fig. 2 (top). The plane chosen for the plots is such that $\theta = 0$, corresponding to a main pole direction. Figure 2 (bottom) plots the radial and axial magnetic field intensity at a fixed radius $r = R_0/2$ along the $z$ direction. One can see how the radial fields decreases close to the PMM edge where an axial peak field is located ($z = \pm 100 \, mm$ here).

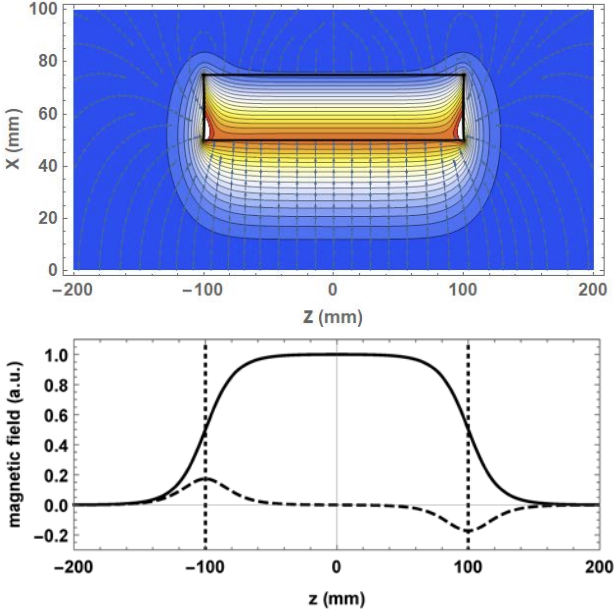

Fig.2. Top: magnetic field lines and magnetic field intensity contour plot of a hexapole PMM with L=200 mm $R_0$ =50 mm and $R_1$ =75 mm. The plane of measurement is y=0, chosen along a main pole. Bottom: (solid) radial magnetic field intensity measured in the same plane (y=0) along the axis z at $r = R_0/2$. (Dashed) axial magnetic field component along z for the same radius. The dotted lines indicates the physical limits of the PMM.

Enge proposed the logistic function to fit the fringing field curve of long multipole accelerator magnets:[4]
$$h(s) = \frac{B_s}{B_0} = \frac{1}{1+e^{S(s)}} \quad (2)$$
where $s$ is the curvilinear abscissa (i.e. curvilinear distance along the reference trajectory) and $S(s) = c_0 + c_1 s + c_2 s^2 + c_3 s^3$, $B_0$ is the constant induction well into the multipole and $B_s$ the local magnetic field taken at $s$. The third order polynomial suits for general multipole magnet shapes including coils and soft iron. Later, Moloney used the Fourier series applied to a quadrupole to calculate the Fringe field [5]. The function used to model the fringe field was there measured and computed numerically. Recently, Muratori found an analytical solution for a multipole's fringe field applicable to numerical simulations [6], but its practical use in simulation appears computationally expensive due to a large number of parameters. In this work, the Fourier multipole expansion is used to derive analytical models applicable to PMM. The previous studies by Enge and Moloney [5,6] are prolongated and a general axial potential function is proposed to model any PMM magnetic field including its fringe fields. The first section presents the Fourier expansion applied to PMM, the magnetic field expansion is proposed for m=1 to 4 multipoles as a function of the axial potential function. The second section is dedicated to the build up of the general axial potential function required in the multipole expansion. The third section proposes a first order model including fringe field and complying with the Maxwell equation. A high accuracy higher order analytical model applicable to simulations is discussed afterward. Finally, a general fit is performed on the unique potential function parameter as a function of the PMM geometry (for m=2,3,4 multipoles). The general fit allows one to generate accurate multipole magnetic field in simulations where the PMM geometry is an input parameter.

## II TAYLOR EXPANSION APPLIED TO MULTIPOLE FRINGE FIELD

In free space, the magnetic field $\vec{B}$ derives from a scalar potential $\Psi$:
$$\vec{B} = -\mu_0 \vec{\nabla} \Psi \quad (3)$$
$\Psi$ satisfies the Laplace equation (expressed below in cylindrical coordinates):
$$\frac{\partial^2 \Psi}{\partial r^2} + \frac{1}{r}\frac{\partial \Psi}{\partial r} + \frac{1}{r^2}\frac{\partial^2 \Psi}{\partial \theta^2} + \frac{\partial^2 \Psi}{\partial z^2} = 0 \quad (4)$$
In the particular case of a multipole whose main pole direction is oriented along the $x$ axis, the Fourier expansion of $\Psi$ reduces to:
$$\Psi(r, \theta, z) = \sum_{n=0}^{\infty} u_n(r,z) \cos n\theta \quad (5)$$
When the function $u_n(r,z)$ is chosen as to separate the variables in the form of series of power of $r$:
$$u_n(r,z) = \sum_{k=0}^{\infty} f_{nk}(z) r^k \quad (6)$$
the substitution of eq. (5) and (6) in eq. (4) and the identification term by term leads to a set of differential equations in which every series component must satisfy eq. (4). After calculations, detailed in [7], the scalar potential expression reduces to:
$$\Psi(r, \theta, z) = \sum_{n=0}^{\infty} \sum_{k=0}^{\infty} \frac{(-1)^k n!}{4^k k!(n+k)!} f_n^{(2k)}(z) r^{n+2k} \cos n\theta \quad (7)$$
where $f_n^{(2k)}$ is the $2k^{th}$ derivative of $f_n(z)$. The planes of symmetry of a multipole of order $N = 2m$ imply supplementary conditions on the possible values of $n$ in eq. (7):
$$n = (1+2l)m \quad (8)$$
Where $l$ is an integer. For instance, for a hexapole n=3,9,15,21, etc. For convenience, the magnetic potential multipolar expansion for m=1 to 4 is proposed in Table 1 and the magnetic field expansion derived with eq. (3) is written in Table 2 in the case of a homogeneous multipole (see III). $B_0$ is the magnetic field intensity at the radius $R_0$, taken at z=0 (see Fig.2).

Table 1: first terms of the Fourier series of the magnetic potential $\Psi$ for common multipoles.

| m | First terms of the magnetic potential multipole expansion |
|---|---|
| 1 | $\Psi_1 = \left(f_1(z)r - \frac{1}{8}f_1^{(2)}(z)r^3 + \frac{1}{192}f_1^{(4)}(z)r^5 - \frac{1}{9216}f_1^{(6)}(z)r^7 + \cdots\right)\cos\theta + (f_3(z)r^5 - \frac{1}{16}f_3^{(2)}(z)r^7 + \cdots)\cos 3\theta + \ldots$ |
| 2 | $\Psi_2 = \left(f_2(z)r^2 - \frac{1}{12}f_2^{(2)}(z)r^4 + \frac{1}{384}f_2^{(4)}(z)r^6 - \frac{1}{23040}f_2^{(6)}r^8 + \cdots\right)\cos 2\theta + (f_6(z)r^6 - \frac{1}{28}f_6^{(2)}(z)r^8 + \cdots)\cos 6\theta + \ldots$ |
| 3 | $\Psi_3 = \left(f_3(z)r^3 - \frac{1}{16}f_3^{(2)}(z)r^5 + \frac{1}{640}f_3^{(4)}(z)r^7 - \frac{1}{46080}f_3^{(6)}r^9 \ldots\right)\cos 3\theta + (f_9(z)r^9 - \frac{1}{40}f_9^{(2)}(z)r^{11} + \cdots)\cos 9\theta + \ldots$ |
| 4 | $\Psi_4 = \left(f_4(z)r^4 - \frac{1}{20}f_4^{(2)}(z)r^6 + \frac{1}{960}f_4^{(4)}(z)r^8 - \frac{1}{80640}f_4^{(6)}r^{10} + \frac{1}{10321920}f_4^{(6)}(z)r^{12} \ldots\right)\cos 4\theta +$ |

$(f_{12}(z)r^{12} - \frac{1}{52}f_{12}^{(2)}(z)r^{14} + \cdots)\cos 12\theta + \ldots$

Table 2: Magnetic field expression derived from the magnetic potential expansion of multipoles for m=1,2,3,4 (see eq. 3 and Table 1).

| m | First terms of Magnetic Field multipole expansion |
|---|---|
| 1 | $B_{r_1} \sim B_0 \left( f_1(z) - \frac{3}{8}f_1^{(2)}(z)r^2 + \frac{5}{192}f_1^{(4)}(z)r^4 - \frac{7}{9216}f_1^{(6)}(z)r^6 + \cdots \right) \cos\theta$ <br> $B_{\theta_1} \sim -B_0 \left( f_1(z) - \frac{1}{8}f_1^{(2)}(z)r^2 + \frac{1}{192}f_1^{(4)}(z)r^4 - \frac{1}{9216}f_1^{(6)}(z)r^6 + \cdots \right) \sin\theta$ <br> $B_{z_1} \sim B_0 \left( f_1^{(1)}(z)r - \frac{1}{8}f_1^{(3)}(z)r^3 + \frac{1}{192}f_1^{(5)}(z)r^5 - \frac{1}{9216}f_1^{(7)}(z)r^7 + \cdots \right) \cos\theta$ |
| 2 | $B_{r_2} \sim \frac{B_0}{R_0} \left( 2f_2(z)r - \frac{1}{3}f_2^{(2)}(z)r^3 + \frac{1}{64}f_2^{(4)}(z)r^5 - \frac{1}{2980}f_2^{(6)}r^7 + \cdots \right) \cos 2\theta$ <br> $B_{\theta_2} \sim -\frac{B_0}{R_0} \left( 2f_2(z)r - \frac{1}{6}f_2^{(2)}(z)r^3 + \frac{1}{192}f_2^{(4)}(z)r^5 - \frac{1}{11520}f_2^{(6)}r^7 + \cdots \right) \sin 2\theta$ <br> $B_{z_2} \sim \frac{B_0}{R_0} \left( f_2^{(1)}(z)r^2 - \frac{1}{12}f_2^{(3)}(z)r^4 + \frac{1}{384}f_2^{(5)}(z)r^6 - \frac{1}{23040}f_2^{(7)}r^8 + \cdots \right) \cos 2\theta$ |
| 3 | $B_{r_3} \sim \frac{B_0}{R_0^2} \left( 3f_3(z)r^2 - \frac{5}{16}f_3^{(2)}(z)r^4 - \frac{7}{640}f_3^{(4)}(z)r^6 - \frac{9}{46080}f_3^{(6)}r^8 + \cdots \right) \cos 3\theta$ <br> $B_{\theta_3} \sim -\frac{B_0}{R_0^2} \left( 3f_3(z)r^2 - \frac{3}{16}f_3^{(2)}(z)r^4 + \frac{3}{640}f_3^{(4)}(z)r^6 - \frac{3}{46080}f_3^{(6)}r^8 + \cdots \right) \sin 3\theta$ <br> $B_{z_3} \sim \frac{B_0}{R_0^2} \left( f_3^{(1)}(z)r^3 - \frac{1}{16}f_3^{(3)}(z)r^5 + \frac{1}{640}f_3^{(5)}(z)r^7 - \frac{1}{46080}f_3^{(7)}r^9 + \cdots \right) \cos 3\theta$ |
| 4 | $B_{r_4} \sim \frac{B_0}{R_0^3} \left( 4f_4(z)r^3 - \frac{6}{20}f_4^{(2)}(z)r^5 + \frac{8}{960}f_4^{(4)}(z)r^7 - \frac{10}{80640}f_4^{(6)}r^9 + \cdots \right) \cos 4\theta$ <br> $B_{\theta_4} \sim -\frac{B_0}{R_0^3} \left( 4f_4(z)r^3 - \frac{4}{20}f_4^{(2)}(z)r^5 + \frac{4}{960}f_4^{(4)}(z)r^7 - \frac{4}{80640}f_4^{(6)}r^9 + \cdots \right) \sin 4\theta$ <br> $B_{z_4} \sim -\frac{B_0}{R_0^3} \left( f_4^{(1)}(z)r^4 - \frac{1}{20}f_4^{(3)}(z)r^6 + \frac{1}{960}f_4^{(5)}(z)r^8 - \frac{1}{80640}f_4^{(7)}r^{10} + \cdots \right) \cos 4\theta$ |

### III DETERMINATION OF THE GENERAL AXIAL POTENTIAL FUNCTION FOR PMM

In the case of a homogeneous multipole field, no azimuthal harmonics are present and $\Psi_m \propto \cos m\theta$. This condition is satisfied for a Halbach structure with a sufficiently large number of magnet per pole and a linear evolution of the magnet axis of magnetization with the magnet number. More precisely, for a multipole of order $m$ with $n$ magnets per pole, a pole sector covers $\frac{\pi}{m}$ and the individual magnet axis of magnetization turns by the angle $\frac{\pi}{n}$. The axial dependence of the potential is then dependent of a single function $f_m(z)$ and its derivatives. For a given multipole of order $m$, when $r \to 0$, $\Psi(r, z, 0) \sim f_m(z) r^m$ as other terms with higher power of $r$ are negligible. In this work, a thorough investigation on the possible function $f_m(z)$ applicable to PMM was performed and lead to the finding of a general analytical potential function fitting any homogeneous PMM magnetic field of arbitrary length $L$:

$$f_{\lambda,L}(z) = \frac{1}{\left(1 + e^{\lambda(z - \frac{L}{2})}\right)\left(1 + e^{-\lambda(z + \frac{L}{2})}\right)} \quad (9)$$

$\lambda$ is a geometrical parameter depending on $R_0$ and $R_1$ only. This function is the product of two logistic functions with a linear term in the exponent. While Enge [4] proposed the logistic function to model the fringe field connected to a sufficiently long internal multipole magnetic field, the present function models the whole magnetic field of the PMM structure including both fringe fields whatever the PMM length L is, with an explicit dependence with $L$. For a given PMM geometry, $\lambda$ is determined by fitting the normalized radial magnetic profile $B_r(r \to 0, \theta = 0, z)$ with $f_{\lambda,L}(z)$. The explicit independence of the parameter $\lambda$ with $L$ has been checked by simulation by varying $L$ on a wide range (from 1 to 500 mm) for several PMM radial geometries $(R_0, R_1)$. A typical evolution of $B_r(z)$ and $B_z(z)$ with $L$ (for $r \to 0$ and $\theta = 0$) is shown in Fig. 3. $B_z(z)$ is connected to $f_{\lambda,L}$ derivative (see section IV) and the same rule applies for the $L$ dependence. The function $f_{\lambda,L}$ can reproduce PMM radial magnetic profile close to the axis even when the fringe field is the dominating effect (e.g. on Fig.3 for $L = 10\ mm$).

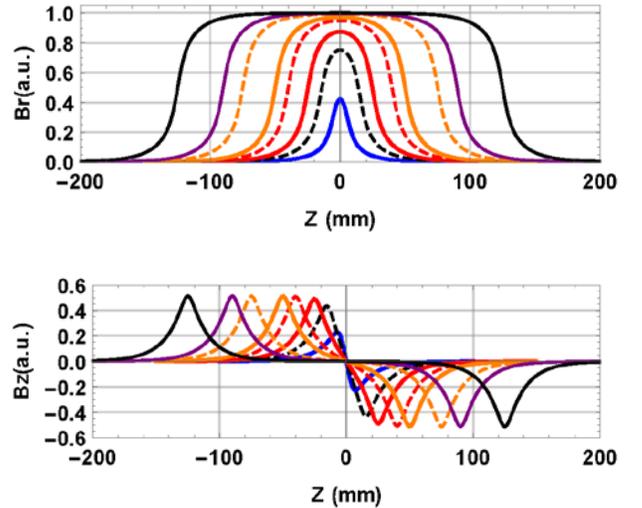

Fig 3: Evolution of the radial (top) and axial (bottom) magnetic field profiles along z as a function of $L$ for a PMM hexapole. The radial intensity is normalized to 1. The different plots correspond to L=10 (blue), 30, 80, 100, 150, 180, 200 mm (black).

### IV FIRST ORDER MULTIPOLE MAGNETIC FIELD MODEL

At the first order of the multipole expansion of Eq. 7, the general expression of the magnetic field takes the form:

$$\begin{cases} B_r = B_0 \left(\frac{r}{R_0}\right)^{m-1} f_{\lambda,L}(z) \cos(m\theta) \\ B_\theta = -B_0 \left(\frac{r}{R_0}\right)^{m-1} f_{\lambda,L}(z) \sin(m\theta) \\ B_z = \frac{B_0 R_0}{m} \left(\frac{r}{R_0}\right)^m f'_{\lambda,L}(z) \cos(m\theta) \end{cases} \quad (10)$$

Where $B_0$ is the magnetic induction at $r = R_0$, $z=0$ and $\theta = 0$. Note that the multipole order $m$ which should naturally appear in the expressions of $B_r$ and $B_\theta$ has been included into the constant $B_0$ to form a more convenient equation. As a consequence, m explicitly appears in the $B_z$ expression as a denominator. $f'_{\lambda,L}$ is the first derivative of $f_{\lambda,L}(z)$ which can be expressed as:

$$f'_{\lambda,L}(z) = -\lambda \left(e^{\lambda(z-\frac{L}{2})} - e^{-\lambda(z+\frac{L}{2})}\right) f_{\lambda,L}^2(z) \quad (11)$$

This first order multipole magnetic field is, by construction, a solution of curl(B)=0 and can be used advantageously in simulations instead of the "hard edge" model as a better approximation of the actual multipole magnetic field. The parameter $\lambda$ is determined by fitting the actual normalized radial magnetic field profile $B_r(r, z, \theta = 0)$ with $f_{\lambda,L}(z)$ at a radius $r \ll R_0$. $B_0$ is deduced from $B_r(r, z = 0, \theta = 0) = B_0(R_0/r)^m$. The derivative $f'_{\lambda,L}$ reproduces the profile $B_z(r, \theta = 0, z)$ that can be seen for instance in the Fig. 2 and 3 bottom. The axial magnetic profile is composed of two peaks, one for each fringing field PMM side. In a real multipole, the axial and radial normalized magnetic profiles shape, taken along z, evolve with the radius, as illustrated in Fig. 4.

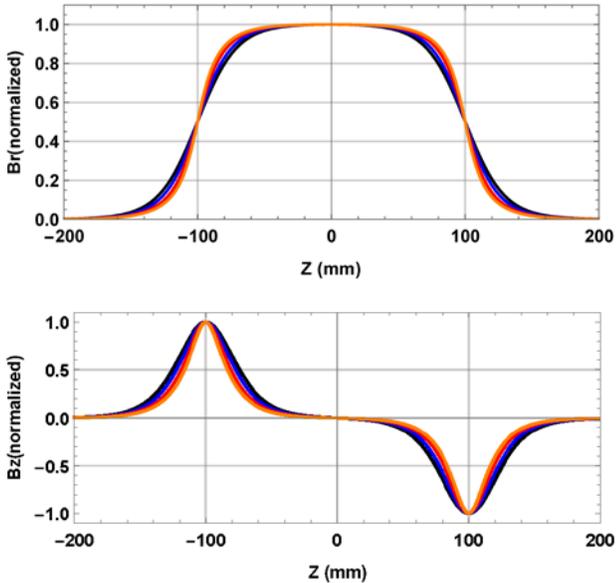

Fig. 4. (Top) Normalized radial (Top) and axial (Bottom) magnetic field profile for r=1 mm (black), 25 mm (blue), 35 mm (red), 45 mm (orange). The geometry is set to $R_0 = 50\ mm$, $R_1 = 75\ mm$, $L = 200\ mm$.

Two $\chi^2$ estimators are built to assess the quality of the first order model with respect to the actual radial and axial magnetic field in the y=0 plane, as a function of the radius $r$:

$$\chi_r^2(r) = \frac{1}{N} \sum_{i=1}^{N} \left(B_r(r, \theta = 0, z_i) - B_0 \left(\frac{r}{R_0}\right)^{m-1} f_{\lambda,L}(z_i)\right)^2 \quad (12)$$

$$\chi_z^2(r) = \frac{1}{N} \sum_{i=1}^{N} \left(B_z(r, \theta = 0, z_i) - \frac{R_0 B_0}{m} \left(\frac{r}{R_0}\right)^m f'_{\lambda,L}(z_i)\right)^2 \quad (13)$$

where $z_i$ ranges from $-2L$ to $2L$ and $N$ is an integer larger than 100. In this study a step of 1 mm step is chosen for the successive $z_i$ values. Next similar functions $\chi_{r,hd}^2$ and $\chi_{z,hd}^2$ are built to compare the actual magnetic field (in the y=0 plane) with the hard-edge model (eq. 1) for the radial and axial magnetic component respectively. Figure 5 plots the evolution of the ratios $\chi_r^2/\chi_{r,hd}^2$ and $\chi_z^2/\chi_{z,hd}^2$ as a function of the normalized radius $r/R_0$ for the multipoles m=1,2,3,4. A ratio lower than one indicates a better fit of the first order model with respect to the hard edge. The benefit of the first order model is obvious for the axial magnetic component for any multipole order m. The first order model is also better for any normalized radius $r/R_0 \leq 0.75$ for any multipole order. The enhancement expands to $r/R_0 \leq 0.9$ since $m \geq 2$. For higher radius ratio, the hard edge is better because close to the magnets, the radial magnetic field shape (versus z) evolves quickly toward a square function (equivalent to setting a high $\lambda$ value of 3-4 in eq. 9), closer to the hard edge model. Usual multipoles applied to accelerators and ion sources have a usable bore of $\sim 0.9 R_0$ and the first order model studied here is clearly a better choice than the hard edge model whatever the multipole order, m, is.

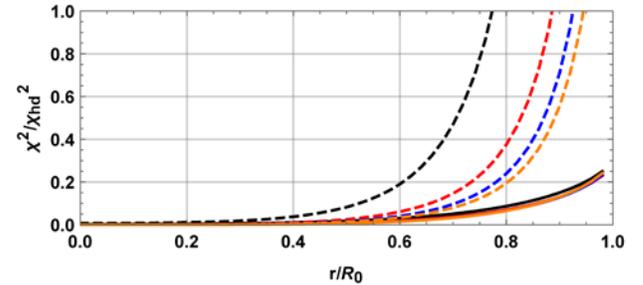

Fig.5. Evolution of the ratio $\chi^2/\chi_{hd}^2$ as a function of the normalized radius for the radial (dashed line) and the axial (solid line) magnetic field components. The different curves stand for m=1(black), m=2 (red), m=3 (blue) and m=4 (orange).

## V ACCURATE ANALYTICAL MODEL OF PMM USING HIGHER ORDER MULTIPOLE EXPANSION

Helped with the axial function $f_{\lambda,L}(z)$ (eq. 9) and the relations in Table 2, a high precision analytical model can be built with the derivatives $f_{\lambda,L}^{(n)}(z)$. The successive derivations of $f_{\lambda,L}^{(1)}$ (eq. 11) can be calculated by hand and cross-checked with a software like Mathematica [8]. A hint to simplify the derivations is to remark that $f_{\lambda,L}^{(1)}$ is the sum of two functions:

$$f_{\lambda,L}^{(1)}(z) = g_{\lambda,\Lambda}(z) + g_{-\lambda,\Lambda}(z) \quad (14)$$

$$g_{\lambda,\Lambda}(z) = -\lambda \Lambda e^{\lambda z} f_{\lambda,\Lambda}^2(z) \quad (15)$$

$$\Lambda = e^{-\frac{|\lambda| L}{2}} \quad (16)$$

$f_{\lambda,L} = f_{\lambda,\Lambda} = \frac{1}{(\Lambda+e^{\lambda z})(\Lambda+e^{-\lambda z})}$ (17)

Using these notations, the successive derivatives of $g_{\lambda,\Lambda}$ express as:

$g_{\lambda,L}^{(n-1)}(z) = (-1)^n \lambda^n \Lambda e^{-(n-2)\lambda z} P_{n,\Lambda}(e^{\lambda z}) f_{\lambda,L}^{n+1}(z), n \geq 2$ (18)

where $P_{n,\Lambda}(x)$ is a polynomial of x. The first 7 polynomials are listed in the Appendix 1 for convenience. A simplification of the derivatives appears when $\lambda L \geq 5$, condition which is commonly fulfilled with the usual PMM geometries ($\lambda > 0.03\ mm^{-1}$, yielding $L \geq 160\ mm$). In this case, the two fringe fields are decoupled and it is possible to express $f_{\lambda,L}^{(1)}$ as the sum of the logistic function derivatives of the two individual fringe fields located at $\pm\frac{L}{2}$:

$f_{\lambda,L}^{(1)}(z) \sim l_\lambda^{(1)}\left(z - \frac{L}{2}\right) + l_{-\lambda}^{(1)}\left(-z + \frac{L}{2}\right)$ (19)

$l_\lambda^{(1)}(z) = \frac{-\lambda e^{\lambda z}}{(1+e^{\lambda z})^2}$ (20)

The general form of $l_\lambda(z)$ derivatives expresses as:

$l_\lambda^{(n)}(z) = \frac{-\lambda^n e^{\lambda z} p_n(e^{\lambda z})}{(1+e^{\lambda z})^{n+1}}$ (21)

but this time the polynomials have a much simpler expression. The first 8 polynomials are listed in the Table 3.

Table 3: $l_\lambda(z)$ derivatives applicable when $\lambda L \geq 5$. Expression of the polynomials $p_n(x)$ for n=1 to 8 appearing in (eq. 19) applicable when $\lambda L \geq 5$.

| |
|---|
| $p_1(x) = 1$ |
| $p_2(x) = 1 - x$ |
| $p_3(x) = 1 - 4x + x^2$ |
| $p_4(x) = 1 - 11x + 11x^2 - x^3$ |
| $p_5(x) = 1 - 26x + 66x^2 - 26x^3 + x^4$ |
| $p_6(x) = 1 - 57x + 302x^2 - 302x^3 + 57x^4 - x^5$ |
| $p_7(x) = 1 - 120x + 1191x^2 - 2416x^3 + 1191x^4 - 120x^5 + x^6$ |
| $p_8(x) = 1 - 247x + 4293x^2 - 15619x^3 + 15619x^4 - 4293x^5 + 247x^6 + x^7$ |

Thanks to the explicit potential function derivatives, the PMM magnetic field multipolar expansion can be calculated to higher orders in $r$. Figure 6 shows how the multipole expansion (m=3 case, hexapole) compares to the actual magnetic field profile at $r = 0.9R_0$, this radius being considered as the practical location of the PMM warm bore location. The different multipole expansion plots stand for the number of r polynomial terms considered: n=1,2,3,4,5. The five first terms are $r^2, r^4, r^6, r^8, r^{10}$ for $B_r$ and $r^3, r^5, r^7, r^9, r^{11}$ for $B_z$. While low order multipolar expansion is sufficient up to $r = 0.5R_0$, higher order is required to grant a high precision reconstruction of the real magnetic field shape up to $r = 0.9R_0$. For the hexapole case, an excellent result is obtained with n=5 up to $\frac{r}{R_0} = 0.9$.

Figure 7 shows the evolution of the $\chi_r^2(r)$ and $\chi_z^2(r)$ (see eq. 12,13) with the normalized radius $r/R_0$ for the hard-edge model (black) and the successive added terms in the multipole expansion model n=1 to 5, as compared to the actual $B_r$ and $B_z$ profiles. One can note that the curves are interleaved, meaning that on some specific radius range, lower order expansion can be more accurate than higher order ones. Anyway, a high quality data fit is obtained with 4 terms at least for both the axial and radial magnetic field.

## VI DEPENDENCE OF $\lambda$ WITH THE MULTIPOLE GEOMETRY

In the former sections, it was demonstrated that a single parameter $\lambda$ (see eq. 9) is sufficient to describe the magnetic field topology of a homogeneous PMM of length $L$, inner and outer radius $R_0$ and $R_1$ respectively (see Fig. 1). $\lambda$ is independent of $L$ and is thus a function the geometrical parameters $R_0$ and $R_1$ only. In this section, the dependence of $\lambda$ with $R_0$ and $R_1$ is studied systematically for the usual multipole cases m=2,3,4. $R_0$ and $R_1$ are varied from 0.1 to 200 mm and 10 to 500 mm respectively. These parameters ranges cover all the practical experimental applications in accelerator and ion source field. Each value of $\lambda$ is determined from the fit of the actual radial magnetic field profile $B_r(r \to 0, \theta = 0, z)$ using eq. 9. The PMM thickness $\Delta R = R_1 - R_0$ is considered as a much practical parameter for the study (see discussion later in the text). The evolution of $\lambda$ with $\Delta R$ is presented in the fig. 8 for various values of $R_0$. The top curves stand for small $R_0$, the lower ones for large $R_0$. When $R_1 \gg R_0$, $\lambda$ is constant which makes sense as the contribution of magnets located at a large distance from the area of interest (i.e. $r \leq R_0$) rapidly tends toward zero. When $R_0$ is small, the fringe field edges are very sharp (high $\lambda$) due to the proximity of the magnets. Indeed, at the magnet surface the $B_r(z)$ profile is close to a "hard edge" shape.

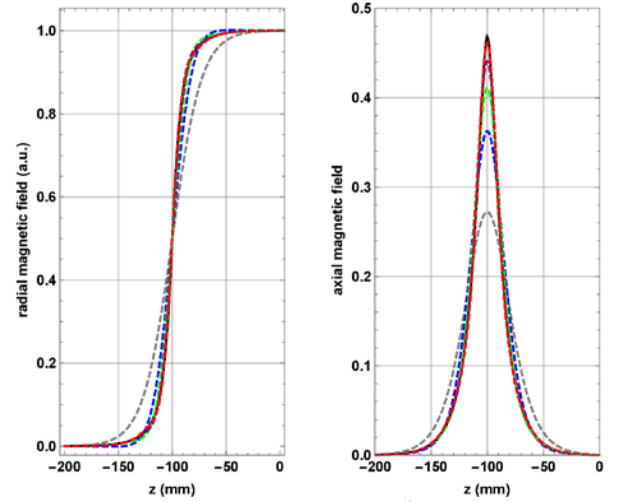

Fig. 6. (Left) Actual $B_r(z)$ profile for $\frac{r}{R_0} = 90\%$ normalized to 1 (black curve) for a hexapole with $L = 200$, $R_0 = 50$ and $R_1 = 75$ mm. The dashed curves are the multipole expansion results for an increasing number of $r$ polynomials terms n=1 (Gray), n=2 (Blue), n=3 (Green), n=4 (Purple), n=5 (Red). (Right) Same plot for the axial magnetic field $B_z(z)$, normalized according to $B_r(z)$, with the same color convention.

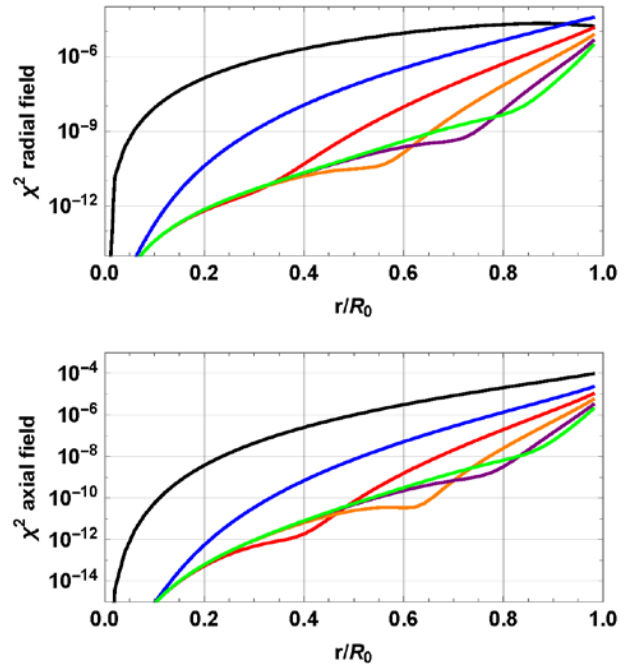

Fig. 7. $\chi^2$ evolution with the normalized radius, comparing the actual radial (top) and the axial (bottom) magnetic field likelihood with the hard edge model (black) and multipole expansion with increasing number of $r$ polynomial terms: blue=1, red=2, orange=3, purple=4, green=5.

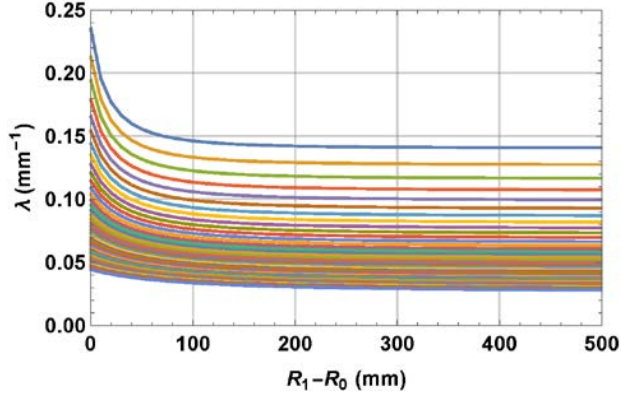

Fig. 8. Evolution of $\lambda(R_0, \Delta R)$ as a function of $\Delta R = R_1 - R_0$ for regularly spaced values of $R_0$ between 0.1 and 200 mm.

Table 4. General fit parameters to calculate $\lambda$ as a function on $R_0$ and $\Delta R$ for the cases m=2,3 and 4.

| fit parameters | m=2 | m=3 | m=4 |
|---|---|---|---|
| $a_\alpha$ | 1.44737 | 2.6803 | 3.58032 |
| $a_\beta$ | 2.58511 | 1.74231 | 2.7946 |
| $b_\beta$ | -2.9123 | -4.68264 | -4.64981 |
| $a_\gamma$ | 3.1224 | 3.13897 | 5.18762 |
| $b_\gamma$ | -2.70186 | -3.71932 | -3.6094 |
| $a_\delta$ | -1.40172e-3 | -8.84823e-4 | -1.12233e-3 |
| $b_\delta$ | 1.27432 | 1.19598 | 1.30156 |
| $c_\delta$ | 18.0588 | 11.0067 | 8.06026 |
| $d_\delta$ | 5.06118 | 2.09334 | -1.04689 |

The function used to fit the dependence of $\lambda$ with $\Delta R$ (for a given $R_0$) is chosen to reproduce both the flat asymptotic behavior for large $\Delta R$ and the exponential evolution when $\Delta R \to 0$:

$$\lambda = \phi(\Delta R) = \alpha + \beta e^{-\gamma \Delta R^\delta} \quad (22)$$

For each $R_0$ value, the fit likelihood is checked with a $\chi^2$ estimator. Finally the evolution of the parameters $\alpha, \beta, \gamma, \delta$ with $R_0$ are fitted with the following empirical functions:

$$\alpha(R_0) = \frac{a_\alpha}{R_0} \quad (23)$$
$$\beta(R_0) = \frac{a_\beta}{R_0 + b_\beta} \quad (24)$$
$$\gamma(R_0) = \frac{a_\gamma}{R_0 + b_\gamma} \quad (25)$$
$$\delta(R_0) = a_\delta R_0 + b_\delta + \frac{c_\delta}{R_0 + d_\delta} \quad (26)$$

Table 4 lists the general fit parameters obtained for the multipoles m=2,3,4. The fits are applicable for $R_0 \geq 5\ mm$ and any $\Delta R$ value. The general fit reproduces the actual $\lambda$ value with a relative precision of ±16%, ±7%, and ±6% respectively for m=2,3,4. When $\Delta R$ is sufficiently large ($\Delta R \geq 150$) the geometrical factor simplifies to $\lambda \sim \frac{a(m)}{R_0}$. The tabulated data allows one to calculate $\lambda$ for a wide range of PMM geometry and to derive analytically the magnetic field using the multipole expansion described in the former section. Fast magnetic field calculations are obtained by tabulating the potential function and its derivatives and by using the multipole expansion of Table 2. The magnetic field intensity $B_0$ is a free parameter of the model.

## APPENDIX

Table 4: polynomials $P_{n,\Lambda}(x)$ appearing in eq. 16.

$P_{n,\Lambda}(x)$ for n=1 to 7

$P_{1,\Lambda}(x)=1$
$P_{2,\Lambda}(x) = \Lambda x^2 - (1 + \Lambda^2)x - 3\Lambda$
$P_{3,\Lambda}(x) = \Lambda^2 x^4 - 4\Lambda(1 + \Lambda^2)x^3 + (1 - 12\Lambda^2 + \Lambda^4)x^2 + \Lambda(1 + \Lambda^2)x + 9\Lambda^2$
$P_{4,\Lambda}(x) = \Lambda^3 x^6 - 11\Lambda^2(1 + \Lambda^2)x^5 + \Lambda(11 - 27\Lambda^2 + 11\Lambda^4)x^4 - (1 - 51\Lambda^2 - 51\Lambda^4 + \Lambda^6)x^3 - \Lambda(1 - 21\Lambda^2 + \Lambda^4)x^2 - 7\Lambda^2(1 + \Lambda^2)x - 27\Lambda^3$
$P_{5,\Lambda}(x) = \Lambda^4 x^8 - 26\Lambda^3(1 + \Lambda^2)x^7 + 6\Lambda^2(11 - 4\Lambda^2 +$

$11\Lambda^4)x^6 + (-26\Lambda + 324\Lambda^3 + 324\Lambda^5 - 26\Lambda^7)x^5 + (1 - 152\Lambda^2 + 612\Lambda^4 - 152\Lambda^6 + \Lambda^8)x^4 + 6\Lambda(1 - 58\Lambda^2 - 58\Lambda^4 + \Lambda^6)x^3 + 2\Lambda^2(9 - 364\Lambda^2 + 9\Lambda^4)x^2 - 26\Lambda^3(1 + \Lambda^2)x + 81\Lambda^4$

$P_{6,\Lambda}(x) = \Lambda^5 x^{10} - 57\Lambda^4(1 + \Lambda^2)x^9 + \Lambda^3(302 + 125\Lambda^2 + 302\Lambda^4)x^8 - 2\Lambda^2(151 - 715\Lambda^2 - 715\Lambda^4 + 151\Lambda^6)x^7 + 3\Lambda(19 - 650\Lambda^2 + 660\Lambda^4 - 650\Lambda^6 + 19\Lambda^8)x^6 - (1 - 395\Lambda^2 + 5140\Lambda^4 + 5140\Lambda^6 - 395\Lambda^8 + \Lambda^{10})x^5 - 7\Lambda(1 - 170\Lambda^2 + 1340\Lambda^4 - 170\Lambda^6 + \Lambda^8)x^4 - 18\Lambda^2(1 - 65\Lambda^2 - 65\Lambda^4 + \Lambda^6)x^3 - \Lambda^3(142 - 4265\Lambda^2 + 142\Lambda^4)x^2 + 347\Lambda^4(1 + \Lambda^2)x - 243\Lambda^5$

$P_{7,\Lambda}(x) = \Lambda^6 x^{12} - 120\Lambda^5(1 + \Lambda^2)x^{11} + \Lambda^4(1191 + 932\Lambda^2 + 1191\Lambda^4)x^{10} - 8\Lambda^3(302 - 587\Lambda^2 - 587\Lambda^4 + 302\Lambda^6)x^9 + \Lambda^2(1191 - 16456\Lambda^2 - 55\Lambda^4 - 16456\Lambda^6 + 1191\Lambda^8)x^8 - 8\Lambda(15 - 1147\Lambda^2 + 5650\Lambda^4 + 5650\Lambda^6 - 1147\Lambda^8 + 15\Lambda^{10})x^7 + (1 - 956\Lambda^2 + 30625\Lambda^4 - 73200\Lambda^6 + 30625\Lambda^8 - 956\Lambda^{10} + \Lambda^{12})x^6 + 8\Lambda(1 - 413\Lambda^2 + 6310\Lambda^4 + 6310\Lambda^6 - 413\Lambda^8 + \Lambda^{10})x^5 + \Lambda^2(31 - 8456\Lambda^2 + 106385\Lambda^4 - 8456\Lambda^6 + 31\Lambda^8)x^4 - 8\Lambda^3(22 - 587\Lambda^2 - 587\Lambda^4 + 22\Lambda^6)x^3 + \Lambda^4(1751 - 21916\Lambda^2 + 1751\Lambda^4)x^2 - 2360\Lambda^5(1 + \Lambda^2)x + 729\Lambda^6$